# The Influence of Cosmic Rays on the Size of the Antarctic Ozone Hole

M. Alvarez-Madrigal[1], J., Pérez-Peraza[2] and V. M. Velasco[2]

[1]Tecnológico de Monterrey, Campus Estado de México, Carretera al lago de Guadalupe, Km. 3.5, Estado de México, CP 52926, MEXICO
[2] Instituto de Geofísica, Universidad Nacional Autónoma de México, México D.F, CP 04510, MEXICO

Correspondence to: M. Alvarez-Madrigal (mmadrigal@itesm.mx)

**Abstract**

The Antarctic region in which severe ozone depletion has taken place is known as the ozone hole. This region has two basic indicators: the area, where the ozone abundance is low (size), and the quantity of ozone mass deficit (depth). The energetic particles that penetrate deeply into the atmosphere and galactic cosmic rays (GCR) modify the ozone abundance in the stratosphere. With this research project, we are looking for evidence of a connection between variations in the cosmic ray flux and variations in the size of the ozone hole. In addition, we are looking for signs of the kind of processes that physically connect GCR fluxes with variations in the stratospheric ozone hole size (OHS) in the Antarctic region. With this goal in mind, we also analyze here the atmospheric temperature (AT) anomalies, which have often been linked with such variations. Using Morlet's wavelet spectral analysis to compute the coherence between two time series, we found that during the analyzed period (1982-2005), there existed a common signal of around 3 and 5 years between the OHS and GCR time series, during September and November, respectively. In both cases, the relationship showed a time-dependent anti-correlation between the two series. On the other hand, for October the analysis showed a time-dependent correlation that occurs around 1.7 years. These results seem to indicate that there exist at least two kinds of modulation processes of GCR fluxes on the OHS that work simultaneously but that change their relative relevance along the timeline.

## 1 Introduction

Since the discovery of the Antarctic ozone hole (OH) by Farman et al. (1985), considerable effort has been focused on observing these ozone losses, on understanding the chemical, dynamical and radiative processes involved, and on predicting the future of polar ozone (World Meteorological Organization (WMO), 2007). It is well known that the main cause of this stratospheric ozone reduction is anthropogenic activity (e. g. Huck et al., 2007), but the influence of precipitating charged particles on the abundance of stratospheric ozone and other atmospheric constituents complicates the interpretation of OH trends (WMO, 2007).

Conventionally, the OHS is calculated from the area contained by total column ozone values less than 220 Dobson Units (DU) (1 DU=0.01 mm. thickness at standard temperature and pressure). The 220-DU value is conventionally used because it corresponds to the strong ozone gradient region (Newman et al., 2004). Figure 1.a presents the monthly averages of OHS trends. In early studies (e. g. Stephenson and Scourfield, 1991; Jackman et al., 1996), the contribution of the precipitating fast charged particles to the ozone mass deficit at the polar latitudes was estimated, but at present it is not clear if such effects are perceptible in large areas around the poles, particularly where ozone abundance is critically low (OHS).

The fast charged particles that influence the atmosphere can be roughly grouped into three types: (1) solar particles, which are mostly protons entering the polar regions and are thus often referred to as solar proton events (SPE); (2) auroral energetic electrons, precipitating in the polar zone and at high latitudes; and (3) galactic cosmic rays, also entering preferentially at high latitudes [WMO, 2007]. Though both SPE and energetic precipitating electrons influence, as we have said above, the polar ozone levels, solar particles primarily deposit their energy in the mesosphere and stratosphere, whereas the auroral energetic electrons primarily deposit their energy in the thermosphere and upper mesosphere.

Since the early studies of Crutzen, et al. (1975) and Heath, et al. (1977), a number of papers have been published documenting the SPE-caused ozone polar changes (Jackman and McPeters, 2004), though the area in which the SPE deplete the ozone has been estimated as a minor one, in relation to the area of chlorine catalyzed depletion (Stephenson and Scourfield, 1991). The GCR continually create odd nitrogen and odd hydrogen constituents in the lower stratosphere and upper troposphere that affect stratospheric ozone abundance, but nowadays it is thought that they play a small role in variations in polar ozone abundance (WMO, 2007).

The understanding of the influence of GCR on the OHS is relevant to the differentiation of the nature of the inducted processes in the atmosphere (natural or anthropogenic) and to the better assessment of the impact of environmental protection policies on indicators of achievement (for example, on the OH extent). It has been well known since the early 1930's that the GCR flux is maximum at polar latitudes and minimum at the equator (e.g., Vallarta, 1933; Lemaitre and Vallarta, 1933; Lu and Sanche, 2001). Depending on their primary energy GCR particles penetrate deep into the atmosphere (even reaching the Earth's surface), altering the ozone abundance at all latitudes: the effects above about 60 degrees are much smaller than at lower latitudes, as particles of energy below ~ 18 GeV are modulated by the Earth's magnetic field. Consequently, their influence on the ozone destruction at polar sites may reach the ceiling, at heights of 10-20 km, where the ionization and dissociation started by GCR is maximal (e. g. Kasatkina and Shumilov 2005).

For this reason, and taking into account that the OHS and GCR time series have similar periodicities in the range of 5-1.7 years (Alvarez-Madrigal, et al., 2008), we consider exploring whether such coincidences do or do not imply a relationship between both phenomena. Therefore, in this study we analyze the monthly mean average of OHS during September, October and November from 1982 to 2005 (Figure 1.a) and the corresponding counts on the neutron monitor at South Pole Station, as a proxy of GCR flux (Figure 1.b). Considering that during the past two decades the global atmosphere was perturbed for several years by two major volcanic explosions - El Chichón volcano in 1982 and Mt. Pinatubo in 1991 (WMO, 2007) -, the expected relation between GCR flux and OHS (if it exists) may be disturbed by such eruptions and probably will be of a non-linear nature.

Taking into account this complex situation, with conditions changing during the analyzed time, with this study we decided to carry out a wavelet analysis, because this statistical tool of analysis allows us to find the frequency components of a time series as a time function and its relative intensity (Hudgins et al. 1993; Torrence and Compo, 1998; Grinsted, Moore and Jevrejera, 2004). It also permits the comparison of two power spectra, in order to show the common frequencies of two series as a time function, by means of the so-called coherence wavelet analysis.

## 2 Data, analysis and results

### 2.1 Data

We worked with monthly data for galactic cosmic ray (GCR) series from 1982 to 2005, taken from the South Pole Station in the Antarctic 90S location (http://neutronm.bartol.udel.edu/~pyle/bri_table.html).

The Ozone Hole (OH) series from 1982 to 2005 were taken from reports of the National Oceanic and Atmospheric Administration (NOAA) Southern Hemisphere Winter Summary 2006 (their figures 5c-5e, available at: http://www.cpc.ncep.noaa.gov/products/stratosphere/winter_bulletins/sh_06).

Tropospheric temperature anomalies (TTA), based on global radiosonde network data (10-15 km) from 1985 to 2005 were taken from NOAA-Air Resources Laboratory (web page: http://cdiac.ornl.gov/trends/temp/contents.htm).

### 2.2 Method of analysis

The simplest technique for investigating periodicities in solar activity is the Fourier transform. Although useful for stationary time series, this method is not appropriate for time series that do not fulfil the steady state condition, such as the time series of the OHS and GCR data.

In order to analyze local variations of power within a single non-stationary time series at multiple periodicities, we applied the wavelet method using the Morlet wavelet (Torrence and Compo, 1998; Grinsted et al., 2004). Wavelet analysis can be used for analyzing localized variations of power within a time series at many different frequencies. The Morlet wavelet consists of a complex exponential $e^{i\omega_0 t/s} e^{-t^2/(2s)^2}$, where $t$ is the time, s is the wavelet scale, and $\omega_0$ is the dimensionless frequency. Here we use $\omega_0 = 6$ in order to satisfy the admissibility condition (Farge 1992). Torrence and Compo (1998) defined the wavelet power $|W_n^X|^2$, where $W_n^X$ is the wavelet transform of a time series X, and n is the time index.

We then estimated the significance level for each scale using only values inside the cone of influence (COI). The COI is the region of the wavelet spectrum in which edge effects become important and is defined here as the e-folding time for the autocorrelation of wavelet power at each scale. This e-folding time is chosen so that the wavelet power for a discontinuity at the edge drops by a factor $e^{-2}$ and ensures that the edge effects are negligible beyond this point (Torrence and Compo, 1998). Wavelet power spectral density (PSD) is calculated for each parameter; the black thin lines mark the interval of 95% confidence, the so-called COI.

The cross wavelet analysis was introduced by Hudgins, et al. (1993). For analysis of the covariance of two time series, we follow Torrence and Compo (1998) and first define the cross wavelet spectrum of two time series OHS and GCR, with wavelet transforms ($W_n^{OHS}$) and ($W_n^{GCR}$), as $W_n^{GCR,OHS} = W_n^{GCR} W_n^{OHS*}$, where (*) denotes complex conjugation. Torrence and Webster (1999) defined the cross-wavelet power as $\left|W_n^{GCR,OHS}\right|^2$. The phase angle of $W_n^{GCR,OHS}$ describes the phase relationship between GCR and OHS in time-frequency space. Unlike the cross wavelet power, which is a measure of the common power, the squared coherency $R_n^2 s$ is used to identify frequency bands within which the two time series are covarying. The wavelet squared coherency is a measure of the intensity of the covariance of the two series in time-frequency space. By definition, the condition $0 \leq R_n^2 \leq 1$ is fulfilled. The sub index (n) indicates the corresponding year for the evaluation.

The statistical significance level of the wavelet coherence is estimated using Monte Carlo methods with red noise to determine the 5% significance level (Torrence and Webster, 1999). If the coherence of two series is high, the arrows in the coherence spectra show the phase between the phenomena. Arrows at 0° (horizontal right) indicate that both phenomena are in phase, and arrows at 180° (horizontal left) indicate that they are in anti-phase; arrows at 90° and 270° (vertical up and down respectively) indicate an out of phase situation. Based on the description given above, we may state that the wavelet coherence spectral analysis is especially useful for highlighting the time and frequency intervals where two phenomena have a strong interaction.

We also include the global spectra, which is an average of the power of each periodicity in both the wavelet and coherence spectra. It allows us to observe at a glance the global periodicities of either the time series or of the coherence analysis (Velasco and Mendoza, 2008). The significance level of the global wavelet spectra is indicated by the dashed curves: they refer to the power of the red noise level at 95% confidence levels that increases with decreasing frequency (Grinsted et al., 2004). It is a way to show the power contribution of each periodicity inside the COI, in which case periodicities are obtained that are on or above the red noise level. The uncertainties of the periodicities of both global wavelet and coherence spectra are obtained from the peak full width at the half maximum of the peak.

We would like to emphasize that the variations in magnitude of the series are not so important as the variations in their periodicities in time, since these give more important information, as is the dependence that exists with other time series (in this case the OHS) in terms of their mutual phase, whether linear or non-linear, whether with correlation or anticorrelation (panels b in figures 2-4). Such information cannot be easily seen through the mere examination of Fig. 1. There are examples, such as the case of solar irradiance, where the magnitude variations in the time series are so small that is designated as "the solar constant". However, its spectral analysis show that it is not a real constant; rather there are important variations in its periodicities on the order of days and years (Kononovich and Mironova, 2006))

## 2.3 Results.

We applied the wavelet coherence analysis to the OHS and the GCR monthly averages, from 1982 to 2005 for September, October, and November respectively; the results appear in Figures 2, 3 and 4. The time series used in the study are shown separately in the upper panel. The wavelet coherence spectrum for each of the series appears in the middle panel of every figure. The global wavelet coherence spectra appear on the rightmost side of the figures. Arrows pointing to the right mean correlation (in phase) and an anticorrelation (in anti-phase) is indicated by a left-pointing arrow. Non-horizontal arrows refer to a more complicated (non-linear) phase difference. In these figures, red indicates high coherence ($R_n^2 = 1$) and blue indicate low coherence ($R_n^2 = 0$); intermediate colours (between red and blue) represent coherence values between 1 and 0, as is shown at the bottom of Figure 2. It must be pointed out that panels a) and b) are totally independent of each other: panel a) is only the plot of two given time series, and panel b) shows the wavelet analysis of these two series.

The coherence between the OHS and GCR for September presents two peaks above the 95% confidence level: common periodicities at 1.3 and 3 years (Figure 2c). Figure 2b shows that the square coherence is in phase and highly significant (0.9 - 1) at 1.3 yeras, between 1982 and 1985. In contrast, the 3-year periodicity is in anti-phase from 1989 to 1997 with a coherence of 0.7, increasing to 0.9 in a complex relationship between 1997 and 2004.

The coherence of OHS with respect to GCR for October shows a common periodicity of 1.7 years above the 95% confidence level (Figure 3c). The square coherence is in phase and highly significant, from 0.9 - 1 between 1982 and 1984. In contrast, this periodicity is in anti-phase from 1988 to 1993 with square coherence of 0.85 falling to 0.65 from 1996 to 2002 (Figure 3b). The rest of the periodicities have low levels of coherence.

Fig. 4a shows the time series of the OHS and GCR for November. It can be seen from the global wavelet coherence spectrum (fig. 4c) that in the 95% confidence region, the most prominent coherences are around periodicities of 1.3, the band 2–4, and 5.5 years. The arrows indicate a tendency to follow an in-phase correlation (at 1.3 years), with an average square coherence around 0.9-1 between 1982 and 1984, falling to 0.7-0.6 in subsequent years. The 2-4 years periodicities are in a complex relationship and the 5.5 year periodicity is in-phase correlation from 1982-1996, changing to complex, from 1997-2005.

The resulting periodicities for the relationship between OHS and GCR during September and October are mostly of a linearly anticorrelated nature. In order to explore the nature of such indirect connections, we also analyzed here stratospheric temperature anomalies (AT) and the GCR flux in the Antarctic region, looking for a possible relation between them. We chose this course of action because temperature anomalies in the Antarctic stratosphere are good explanatory variables for interannual variability of the OHS anomalies (Huck, et al. 2007), although results have also been obtained on the basis of annual indicators.

Thus we obtain here the coherence between the power spectra of GCR and AT, at 15 km height, using as indicators the average values of the AT anomalies and GCR fluxes during the analyzed months. The results are shown in Figure 5. In this figure we can observe a square coherence level up to 0.85 at a frequency of around 1.7 years, showing a linear correlation during the period 1987-1993 and a tendency to an anticorrelated relationship, during 1997-2000 (of about 0.6 of square coherence). The relation is complex for the 3-year periodicity in the period (1999-2004).

It should be noted that the main interaction between GCR and AT occurs at the 1.7-year frequency, the same as that between GCR and OHS during the months of October, but GCR are in phase with AT whereas they are in anti-phase with OHS, at least before 1993. After 1995 there is an anticorrelation tendency in both cases. The only thing that we can infer from this is that during the month of October, there is a kind of "resonance cavity" in the 1.7-year frequency at a level of 15 Km, which influences the OHS with a combination of both phase and-anti-phase effects. Most likely this resonance is more relevant in October, when the OHS is full developed, in contrast with September when the OH is opening; meanwhile, in November there is a tendency toward occlusion (e. g. Alvarez-Madrigal and Pérez-Peraza, 2005).

It should be noted that during September and October, there is a clear change in the linear relationships (from positive to negative) around the time interval in which changes in the atmospheric chemical composition were inducted, as a consequence of the two mayor volcano eruptions (1982, 1991). Such a kind of change is not seen in November. At the moment we can only speculate that the combined effect of occlusion with the direction and intensity of winds during November made the difference with respect to September and October.

## 3 Discussion

The fact that the nature of the GCR and OHS connection evolves in time in a linear way (in phase, anti-phase or out of phase) led us to think at first about the joint action of at least two kinds of mechanisms of influence of GCR on the OHS: a mechanism to explain the in-phase correlations and another to explain the anti-phase correlation. The most commonly accepted mechanisms of influence between GCR and stratospheric ozone abundance are (a) the ozone destruction related to the dissociation of chlorofluorocarbons (CFC) through the capture of electrons produced by cosmic rays localized in polar stratospheric cloud ice (Lu and Sanche 2001) and (b) the ozone destruction induced by physical-chemical processes, via the created odd nitrogen and odd hydrogen through GCR impacts (e.g. WMO, 2007).

Mechanism (a) implies that when the GCR flux is maximum, the ozone destruction is maximum and the relation with the OHS will be linear positive; high GCR flux implies high ozone destruction and, consequently, an area of more ozone depletion. Mechanism (b) implies an indirect connection, and the implied relation may be non-linear, but, in any case, of a positive nature. The combined action of these mechanisms may be resumed in three scenarios: i) mechanism (a) is predominant; ii) mechanism (b) is prevalent; and iii) neither one is dominant. In scenario i) the expected relationship between GCR fluxes and OHS is linear positive. For scenarios ii) and iii) a non-linear relationship is expected; however, we always expect a positive relation. In other words, based on these mechanisms, if GCR flux increases, we expect an increase in the OHS. The combined action of mechanisms (a) and (b) may explain the obtained complex and the linear positive relations but cannot explain the anticorrelations detected.

Anticorrelation cases imply that an increase in GCR flux provokes a reduction of the OHS. Currently, we do not have an explanation for this fact. We note that such anticorrelation more often occurs during periods of high solar activity, when the solar magnetic field undergoes changes in polarity (Figs. 2 and 3) and GCR is strongly reduced. However, no matter what kind of modulation exists, that could lead to an anticorrelation effect; this scenario is not relevant for the month of November (Fig. 4). The idea could be explored that GCR particles may dissociate the stratospheric molecular oxygen, while increasing the atomic oxygen levels, and thus also increasing the rate of ozone formation. As a result, there would be a reduction of the area where the stratospheric ozone is critically low (OHS). In fact, since the ozone molecule is similar to the oxygen molecule, it is then expected that GCR particles will dissociate both molecules (like it does with solar ultraviolet radiation), reaching a dynamical equilibrium with the new rates of creation and destruction. However, this idea needs an explanation: why is the projected level of ozone more abundant, and what are the conditions needed for these processes to take place. To explain these relationships, further research is needed to specify the acting mechanisms and to clarify the physical conditions needed for them to occur.

## 4   Conclusions

The results shown in Figure 2 imply common periodicities around 1.3 and 3-years between OHS and GCR during the month of September. Figure 4 shows defined 5.5- and 1.3-year periodicities and an irregular band in the range of 2 to 4 years during the month of November. In both cases the relationship is one of anticorrelation with a tendency toward a complex one in intermittent periods after 1995. Figure 3 shows a predominant in-phase common frequency at 1.7 years. All of these periodicities are in the range of the mid-term periodicities that appear in analyses of solar variability (Mendoza, Velasco and Valdés-Galicia, 2006) and cosmic ray flux (Valdés-Galicia and Velasco, 2005). In particular, the 3-year frequency also appears in studies related to the effect of solar activity and cosmic rays on terrestrial parameters, particularly the Atlantic Multidecadal Oscillation (AMO), which in turn is associated with the North Atlantic sea surface temperature and with droughts in North Europe and North America (e.g. Velasco and Mendoza, 2008).

The mid-term periodicities in the solar activity, which modulates the GCR propagation in the heliosphere, are often attributed to the effect of solar magnetic flux transport which causes the horizontal dipole to decay on the meridional flow. Most probably they are generated in the base of the convective zone namely the *tachocline* (e.g., Howe et al., 2000; Wang and Sheeley, 2003; Wang, 2004, Livshits and Obridko, 2006; Obridko and Shelting, 2007). Besides, according to Benevolenskaya (1995) the solar magnetic cycle consists of two cycles: low-frequency (22-years) and *high-frequency* (1.5-2 yrs.).

Based on our previous results, we have also concluded that at least two classes of physical mechanisms are acting simultaneously, which explains the positive and the anticorrelation relationships found in this study. The positive relation between GCR and OHS is consistent with the mechanisms that are commonly used to explain the ozone depleted by GCR: the dissociation of the CFCs by means of the electrons produced by GCR collisions with the media and via the odd nitrogen and odd hydrogen induced by GCR impacts on the atmosphere. However, we cannot explain the anticorrelation case based only on the most common mechanism of influence of fast particles on the stratospheric ozone. Further research is needed to establish a connection between the dominant physical processes and the sense of the relationship GCR-OHS found in this coherence index analysis. Furthermore, from our analysis we cannot drawn any conclusion about the joint effect of GCR flux variations and AT variations on the behaviour of the OHS.

It should be emphasized that GCR are modulated not only by solar activity but also by the heliospheric magnetic field. In addition to the well-known 11-year periodicities, the GCR time series show others at 1.3, 1.7, 3, 5.5, and 7 years (e. g., Mavromichalaki, et al., 2003; Valdés-Galicia et al., 2005). Using Beryllium 10 and Carbon 14 as proxies for GCR, periodicities of 30 and 60 years have been found, as well as secular periodicities (longer than 100 years) and super-secular periodicities (longer than 500 years) (Velasco et al., 2008). Short periodicities in the range of months have been also reported in solar flare activity (Rieger et al., 1984), and we know that GCR are modulated by solar activity.

Now our main point is that data on OHS only cover 23 years, and we would need data covering more solar cycles to determine the nature of the 11-year periodicity within the framework of our results. As a matter of fact, an examination of panels b) of figs 2-4, shows that outside of the COI, there is a systematic periodicity of around 11 years throughout the analyzed period (1982-2005), but we have not included it in our calculations because is outside the confidence cone. It is quite likely that there exist periodicities more important than those found here, but unfortunately we cannot determine them using the limited data set of OHS. In all probability, other periodicities of longer duration may have a greater influence on the OHS than those established in this study. Nevertheless, the results reported in this paper lead to the assumption that the stratospheric layer may be considered to act as a "resonance cavity" of the GCR variations at frequencies of 1.3, 1.7, 3 and 5.5 years.

**References**
Alvarez-Madrigal, M., and Pérez-Peraza J.: Analysis of the evolution of the Antarctic ozone hole size, J. Geophys.Res., 110, D02107, doi:10.1029/2004JD004944, 2005.
Alvarez-Madrigal M., and Pérez-Peraza J.: Differences in the monthly evolution of the Antarctic ozone hole size, Atmósfera, 20(2), 215-221, 2007.
Alvarez-Madrigal M., Pérez-Peraza J. and Velasco V. M.: On a plausible relationship between Cosmic Rays and the ozone hole size, Proceedings of 30 ICRC, Mérida, México, 1,SH, 789-792, 2008.
Benevolenskaya, E.E.: Double Magnetic Cycle of Solar Activity Solar Phys., **161**, 1-8, 1995.
Bodeker, G. E., Shiona, H. and Eskes H.: Indicators of Antarctic ozone depletion, Atmos. Chem. Phys., 5, 2603–2615, 2005.
Crutzen, P.J., Isaksen, I.S.A., and Reid, G.C.: Solar proton events: Stratospheric sources of nitric oxide, Science, 189 (4201), 457-459, 1975.
Farge, M., Wavelet transforms and their applications to turbulence. Annual Review of Fluid Mechanics 24, 395–457, 1992..
Farman, J.C., Gardiner B.G, and Shanklin, J.D.: Large losses of total ozone in Antarctica reveal seasonal ClOx/NOx interaction, Nature, 315, 207-210, 1985.
Grinsted, A., Moore, J., and Jevrejera, S.: Application of the cross wavelet transform and wavelet coherence to geophysical time series, non-linear processes. Geophys., 11, 561-566, 2004.
Heath, D.F., Krueger, A.J., and Crutzen, P.J.: Solar proton event: Influence on stratospheric ozone, Science, 197 (4306), 886-889, 1977.
Howe, R., Christensen-Dalsgaard, J., Hill, F., Komm, R. M. Larsen, J. Schou, M. J. Thompson, and J. Toomre.: Dynamic Variations at the Base of the Solar Convection Zone, Science **287**, 2456, 2000.
Huck, P. E., A. J., McDonald, G. E. Bodeker, and Struthers H.: Interannual variability in Antarctic ozone depletion controlled by planetary waves and polar temperature, Geophys. Res. Lett., 32, L13819, doi:10.1029/2005GL022943, 2005.
Hudgins L., Friehe C. A., and Mayer M.E.: Wavelet transforms and atmospheric turbulence, Phys. Rev. Lett., 71, 3279-3282, 1993.

Jackman, C. H., E. L. Fleming, S. Chandra, D. B. Considine, and Rosenfield J.E.: Past, present, and future modeled ozone trends with comparisons to observed trends, J. Geophys Res., 101, 28,753-28,767, 1996.

Jackman, C.H., and McPeters R. D.: The effect of solar proton events on ozone and other constituents, in solar variability and its effects on climate, edited by J.M. Pap, P.A. Fox, and C. Frohlich, AGU Monograph 141, 305-319, Washington, D.C., 2004.

Kasatkina, E. A., and Shumilov O. I.: Cosmic ray-induced stratospheric aerosols: A possible connection to polar ozone depletions, Annales Geophysicae, 23, 675-679, 2005, ( SRef-ID: 1432-0576/ag/2005-23-675).

Kononovich, E.V. and Mironova, I.V.: The wolf number and total irradiance variations during 21 and 23 solar cycles, Astronomical and Astrophysical Transactions I 25, 341-345, 2006, (DOI : 10.1080/10556790601106324).

Lemaitre, G and Vallarta, M.S.: On Compton's latitude effect of cosmic radiation, Phys. Rev. 43(2), 87-91, 1933.

Livshits, I.M. and Obridko, V.N.: Variation of the Dipole Magnetic Moment of the Sun during an Activity Cycle, Astronomicheski Zhurnal **83(11),** 1031-1041, 2006.

Lu Q.-B. and Sanche L.: Effects of Cosmic Rays on Atmospheric Chlorofluorocarbon Dissociation and Ozone Depletion, Phys. Rev. Lett., 87,7,078501(4), doi:10.1103/PhysRevLett.87.078501, 2001.

Mavromichalaki H., et al., Low and high frequency spectral behavior of cosmic ray intensity for period 1953-1996, Annales Geophysicae,21, 1681-1689, 2003.

Mendoza, B., Velasco, V. and Valdés-Galicia, J.F., Mid-Term Periodicities in the Solar magnetic Flux// Solar Physics ISSN: 0038-0938.: 233, 319-330, 2006.

Newman P. A., Kawa R., and Nash E. R.: On the size of the Antarctic ozone hole, Geophys. Res. Lett.,31,L21104, doi:10.10292004GL020596, 2004.

Obridko, V.N. and Shelting, B.D.: Occurrence of the 1.3-year periodicity in the large-scale solar magnetic field for 8 solar cycles, Adv. In Space Res., 40, 1006-1014, 2007.

Rieger, E., Share G. H., Forrest D. J., Kanbach G., Reppin C., and Chupp E. L.: A 154-day periodicity in the occurrence of hard solar flares?, *Nature*, 312, 623-625, 1984.

Stephenson J. A. E, and Scourfield M. W. J.: The importance of energetic solar protons in ozone depletion, Nature, 352,137-139, 1991.

Torrence, C. and Compo, G.: A practical guide to wavelet analysis. Bull. Amer. Meteor. Soc., 79, 61-78, 1998.

Valdés-Galicia,J:F:, Velasco, V and Mendoza, B.; Mid Term Cosmic Ray Quasi Periodicities And Solar Magnetic Activity Manifestations// 29th International Cosmic Ray Conference Pune, India, SH24, 101-104, 2005

Vallarta, M.S., The interpretation of the azimuthal effect of cosmic radiation, Phys. Rev. 44(1), 1-3, 1933.

Velasco, V. , Mendoza, B. and Valdés-Galicia J.F., The 120-year solar cycle of the cosmogenic isotopes, Proc. 30th ICRC, 1, 553-556, 2008.

Velasco V. M., and Mendoza B.: Assessing the relationship between solar activity and some large scale climatic phenomena, Adv. Space. Res., 42, 866–878, 2008.

Wang, Y.-M.: The Sun's large-scale magnetic field and its long-term evolution, Solar Phys., 224, 21-35, 2004.

Wang Y.-M. and Sheeley Jr, On the fluctuating component of the sun´s large scale magnetic field, ApJ., 590(2), 1111-1120, 2003.

World Meteorological Organization: Scientific Assessment of Ozone Depletion: 2006, WMO Global Ozone Res. and Monit. Project Rep. 50, Geneva, 2007.

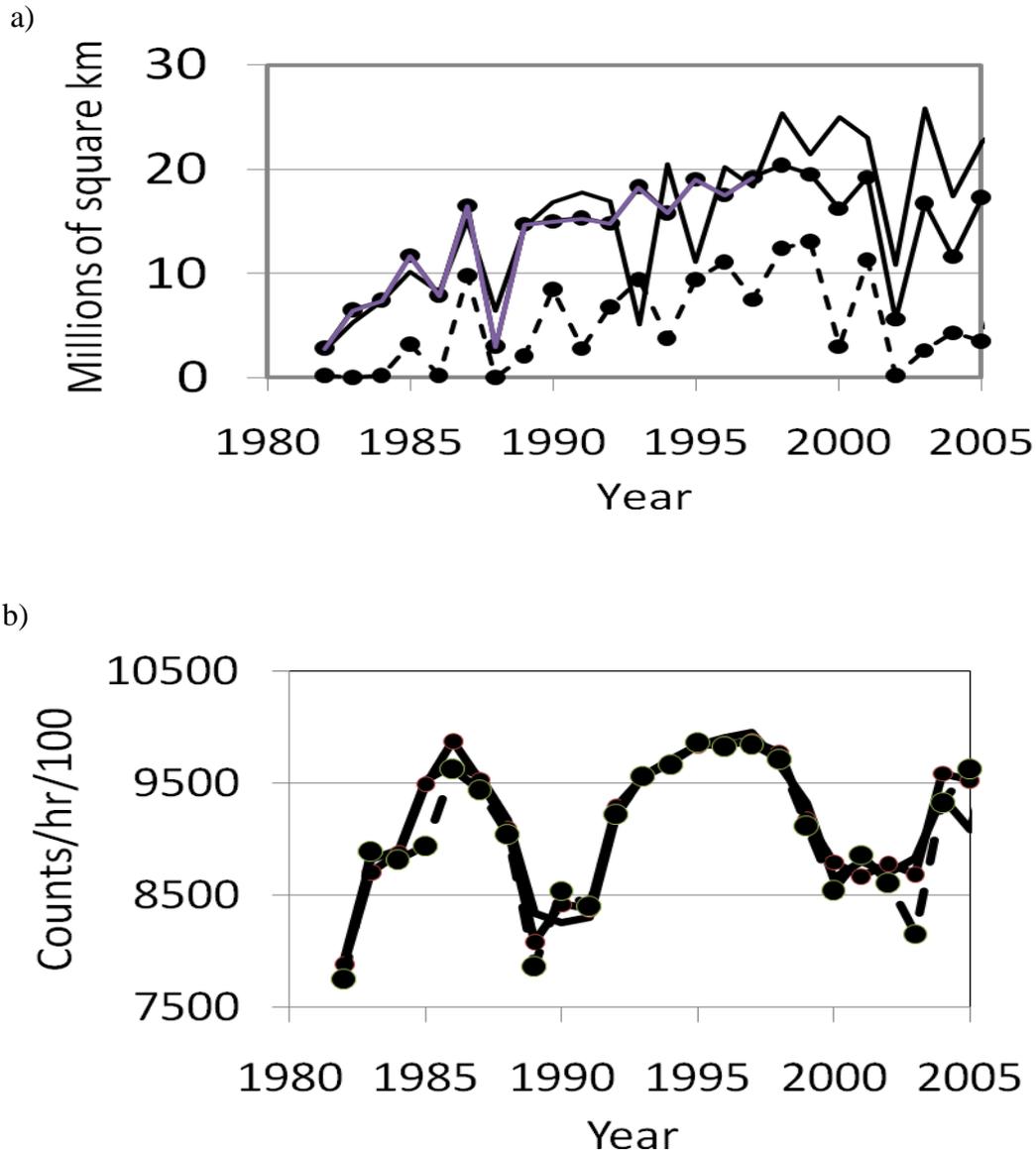

Figure 1. In the upper panel (a) the monthly mean average of the Antarctic ozone hole size is shown from 1982 to 2005, as reported by NOAA. In the bottom panel (b) the monthly mean average of the counts from the neutron monitor detectors at the South Pole station is shown. The continuous line illustrates the September average, the dotted line shows October, and the dashed pointed line shows November data.

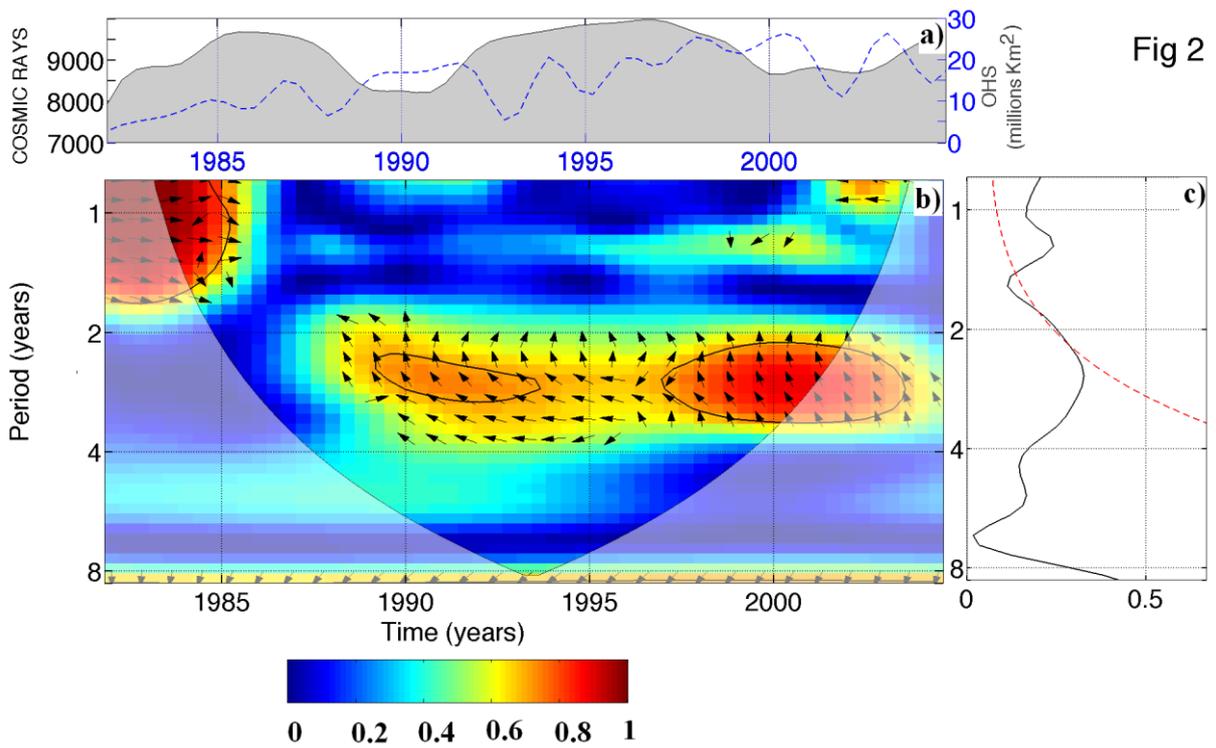

Figure 2. Square coherence between OHS and GCR power spectra. Data corresponding to September (1982 to 2005). In the upper panel (a) the dotted line corresponds to OHS data series, in millions of square kilometers and the continuous line represents the GCR flux, in hundreds of counts/hour. The bottom panel (b) shows the comparison of the power spectra of both data series. In the right panel (c) the continuous line represents the global wavelet spectrum, and the red pointed line shows their 95% confidence limit. At the bottom is the color code.

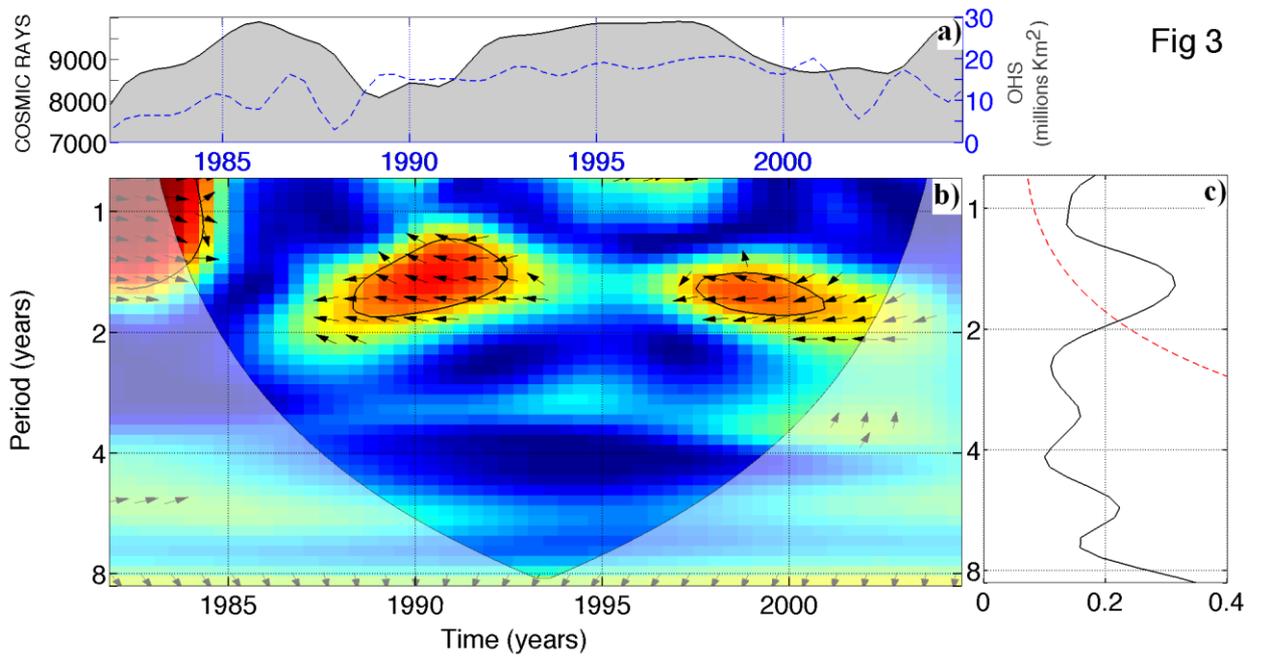

Figure 3. Square coherence between OHS and GCR power spectra. Data corresponding to October (1982 to 2005). In the upper panel (a) the dotted line corresponds to OHS data series, in millions of square kilometers, and the continuous line represents the GCR flux, in hundreds of counts/hour. The bottom panel (b) shows the comparison of the power spectra of both data series. In right panel (c) the continuous line represents the global wavelet spectrum, and the red pointed line shows their 95% confidence limit.

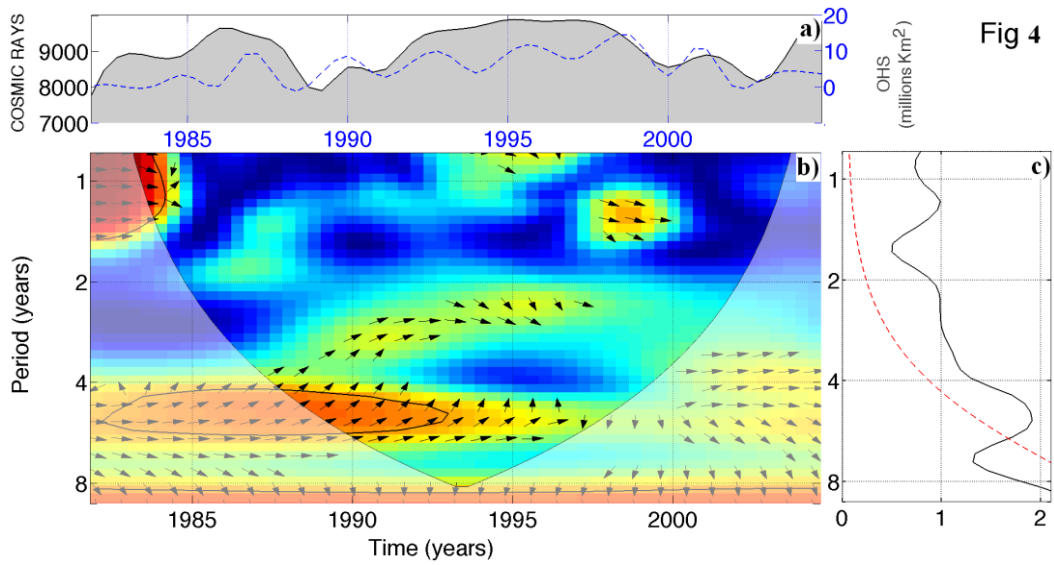

Figure 4. Square coherence between OHS and GCR power spectra. Data corresponding to November (1982 to 2005). In the upper panel (a) the dotted line corresponds to OHS data series, in millions of square kilometers, and the continuous line represents the GCR flux, in hundreds of counts/hour. The bottom panel (b) shows the comparison of the power spectra of both data series. In right panel (c) the continuous line represents the global wavelet spectrum, and the red pointed line shows their 95% confidence limit.

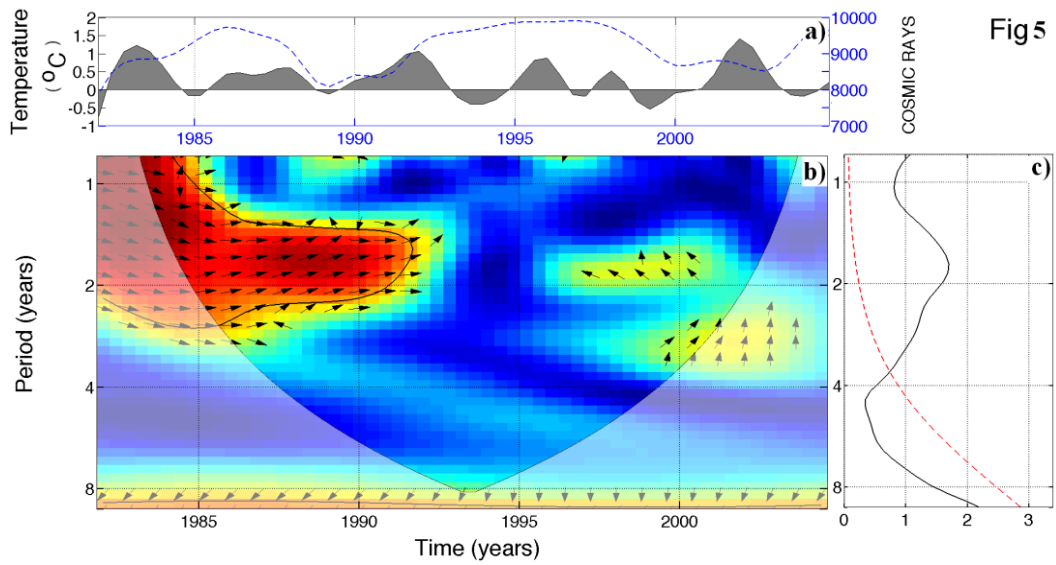

Figure 5. Square coherence between AT and GCR power spectra. Data corresponding to September-November averages (1982 to 2005). In the upper panel (a) the continuous line corresponds to AT data series in Kelvin degrees, and the dotted line represents the GCR flux, in hundreds of counts/hour. The bottom panel (b) shows the comparison of the power spectra of both data series. In right panel (c) the continuous line represents the global wavelet spectrum, and the red pointed line shows their 95% confidence limit.